\documentclass[runningheads]{llncs}
\usepackage{graphicx}
\usepackage{bm}
\usepackage{todonotes}
\usepackage{mathtools}
\usepackage{multirow}
\usepackage{hyperref}
\usepackage{subcaption}
\usepackage{booktabs}
\usepackage{amsmath}
\usepackage{amssymb}
\usepackage[misc]{ifsym}
\usepackage[T1]{fontenc}
\usepackage{diagbox}
\usepackage{academicons}
\usepackage{orcidlink}

\usetikzlibrary{calc}

\begin{document}
\title{More From Less: Self-Supervised Knowledge Distillation for Routine Histopathology Data}
\titlerunning{More From Less: Self-Supervised Knowledge Distillation}

\author{\orcidlink{0009-0003-3667-2001} Lucas Farndale\inst{1,2,3,4}\textsuperscript{(\Letter)} \and
\orcidlink{0000-0003-4898-040X} Robert Insall\inst{1,2} \and
\orcidlink{0000-0002-2318-1460} Ke Yuan\inst{1,2,3}}

\authorrunning{L. Farndale et al.}

\institute{School of Cancer Sciences, University of Glasgow, Glasgow, Scotland, UK \and
Cancer Research UK Beatson Institute, Glasgow, Scotland, UK \and
School of Computing Science, University of Glasgow, Glasgow, Scotland, UK \and
School of Mathematics and Statistics, University of Glasgow, Scotland, UK\\
\email{\{lucas.farndale, robert.insall, ke.yuan\}@glasgow.ac.uk}}
\maketitle
\begin{abstract}
    Medical imaging technologies are generating increasingly large amounts of high-quality, information-dense data. Despite the progress, practical use of advanced imaging technologies for research and diagnosis remains limited by cost and availability, so more information-sparse data such as H\&E stains are relied on in practice. The study of diseased tissue would greatly benefit from methods which can leverage these information-dense data to extract more value from routine, information-sparse data. Using self-supervised learning (SSL), we demonstrate that it is possible to distil knowledge during training from information-dense data into models which only require information-sparse data for inference. This improves downstream classification accuracy on information-sparse data, making it comparable with the fully-supervised baseline. We find substantial effects on the learned representations, and pairing with relevant data can be used to extract desirable features without the arduous process of manual labelling. This approach enables the design of models which require only routine images, but contain insights from state-of-the-art data, allowing better use of the available resources.
    
\end{abstract}

\section{Introduction}

The complexity and amount of information generated by medical imaging technologies is constantly increasing. Developments such as spatial -omics, multiplex immunohistochemistry and super-resolution microscopy are continuously enabling greater insights into mechanisms of disease, but adoption of such technologies is prohibitively expensive for large cohorts or routine use. It is therefore highly desirable to develop methods which distil knowledge from these exceptionally dense and information-rich data into more accessible and affordable routine imaging models, so clinicians and researchers can obtain the most diagnostic information possible from the data available to them.

Typically, knowledge distillation focuses on distilling from a (possibly pretrained) larger teacher model into a smaller student model \cite{gou2021knowledge}. This is usually achieved using a self-supervised joint-embedding architecture, where two models are trained as parallel branches to output the same representations \cite{gou2021knowledge}, so the smaller model can be more easily deployed in practice on the same dataset without sacrificing accuracy. This approach is ideal for digital pathology in which complete images are impractically large and often initially viewed at low magnification. Both knowledge distillation \cite{javed2023knowledge} and SSL \cite{quiros2022self} have been shown to improve performance on histopathology imaging tasks, including when used in tandem \cite{dipalma2021resolution}.

Existing approaches usually require that the data sources used in training are both available during inference, which is severely limiting where at least one source of data is not available, is more expensive, or is more computationally demanding. There are a limited number of SSL architectures which can train two encoders with different inputs concurrently, such as CLIP \cite{radford2021learning}, VSE++ \cite{faghri2017vse++}, Barlow Twins \cite{zbontar2021barlow}, SimCLR \cite{chen2020simple} and VICReg \cite{bardes2021vicreg}, which has been shown to outperform VSE++ and Barlow Twins on multi-modal data \cite{bardes2021vicreg}.

In this work, we focus on knowledge distillation from models of \emph{information-dense} datasets (datasets rich in accessible information, e.g. high-resolution pathology images) into a models of \emph{information-sparse} datasets (datasets containing little or obfuscated information, e.g. low-resolution images). We make the following contributions:
\begin{itemize}
    \item We find that knowledge distillation significantly improves downstream classification accuracy on information-sparse data, comparable to a supervised baseline;
    \item We show that this training process results in measurably different representations compared to standard self-supervised and supervised training, and that the paired data significantly changes the areas of the image focused on by the model;
    \item We show the clinical utility of this method, by presenting a use-case where pan-cytokeratin (pan-CK) stained immunofluorescence (IF) imaging is used to train a better performing model for brightfield hematoxylin and eosin (H\&E) stains.
\end{itemize}

\begin{figure}[t]
    \centering
    \includegraphics[width=\textwidth]{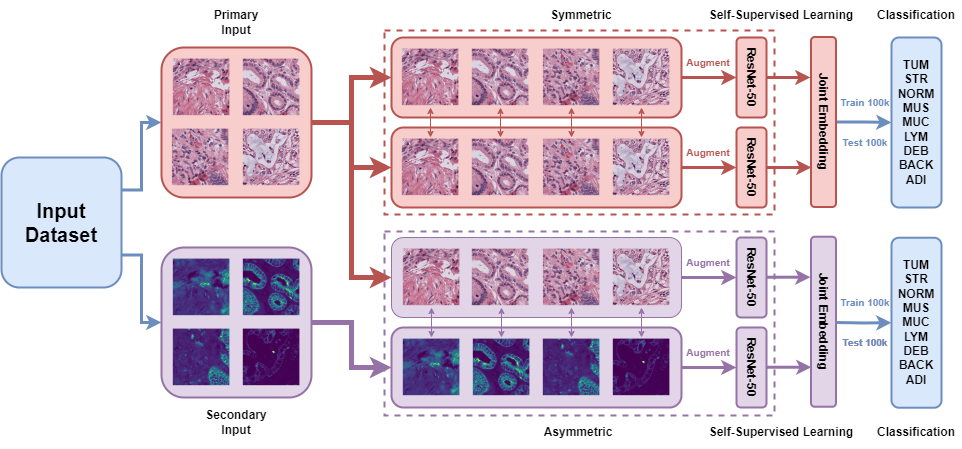}
    \caption{Knowledge distillation between information-dense and information-sparse data. The upper branch shows a symmetric model, while the lower branch shows an asymmetric model. Images are preprocessed (augmentations detailed in Table \ref{tab:augmentations}), paired and passed to the self-suprevised model for training. The trained encoders are then used for downstream classification tasks to assess performance.}
    \label{fig:dataset}
\end{figure}

\section{Methods}

\subsection{Experimental Design}

\noindent We use the self-supervised methods VICReg \cite{bardes2021vicreg} and SimCLR \cite{chen2020simple} to distil knowledge from information-dense data to information-sparse data. As shown in Fig.~\ref{fig:dataset}, the model takes pairs of images as input, one for each branch. We refer to a pair consisting of two copies of the same image as a \emph{symmetric} pair, and a pair consisting of distinct images as an \emph{asymmetric} pair. For example, the primary input in an asymmetric pair might always be a lower resolution than the secondary input, while in a symmetric pair, both inputs would be the same resolution.

\subsubsection{H\&E/IF Distillation}

We first demonstrate the efficacy of asymmetric training with a clinically-relevant use-case, with models trained on the SHIFT dataset \cite{burlingame2020shift}, which contains $256\times256$px patches from H\&E stains, which are washed, restained with pan-CK and re-imaged. Pan-CK stains tumour cells, making it easier to differentiate tissue types. The trained models are evaluated on the standard tissue classification task using the NCT dataset \cite{kather_dataset}, with label distributions shown in Table \ref{tab:class_balance}.

\subsubsection{Contextual Distillation}

Patches are usually created of an arbitrary, small size, making it difficult to detect patterns that may be obvious in larger patches. We show the effects of pairing inputs which can utilise surrounding contextual information, such as the area surrounding the border of a patch. We use the NCT dataset \cite{kather_dataset} to create two examples for original patches paired with: i) patches centre-cropped to $112\times112$px and zero-padded back to $224\times224$px, ii) patches downsampled to $7\times 7$px and resized to $224\times224$px. Padding/resizing is to ensure consistency between encoders in asymmetric/symmetric models.

\subsubsection{Nuclear Segmentation Distillation}

To enable finer-grained analysis with a toy example, we synthetically construct an information-sparse dataset by using HoVer-Net \cite{graham2019hover} to create a nuclear segmentation mask for each image, coloured by their predicted cell types (Background, Neoplastic, Inflammatory, Connective, Dead, Non-Neoplastic Epithelial). We repeat these 1D masks 3 times along channel 3 to ensure consistency of input size between models. Synthetic masks were used due to the limited number of manually annotated masks available, as SSL typically requires large datasets, as demonstrated in Table \ref{tab:dataset_size}, and due to the lack of datasets with both nuclear segmentations and tissue type labels. These images contain significantly less relevant information for tissue type classification, as the majority of fine-grained morphological information in the image is lost, and the HoVer-Net segmentations are imperfect. The masks do not introduce additional information, as by definition all information which is present in these nuclear masks is present in the original images. In Table \ref{tab:conic_results} we show that the method is robust to the use of manually annotated or synthetic data for evaluation.

We evaluate models' performance on the same tissue classification task as above for both the H\&E patches and the masks, and we also evaluate performance on predicting the most common cell type in each image, to illustrate how learned representations transfer between tasks. For comparison, we include supervised models trained on tissue or cell classification before freezing weights, re-training the classifier head and evaluating on the other task.

\subsection{Training}

In our experiments\footnote{Code is available at https://anonymous.4open.science/r/More-From-Less.}, all models used a ResNet-50 encoder \cite{he2016deep} with output size 2048, and expanders were composed of three dense layers with ReLU activations and output size 8192. Encoder ablations are detailed in Table \ref{fig:encablations}. Models are trained using Tensorflow 2.9.1 from a random initialisation for 100 epochs, with a batch size of 128. An Adam optimiser \cite{kingma2014adam} with a warm-up cosine learning rate \cite{goyal2017accurate} was used, with a maximum learning rate of $10^{-4}$, and warm-up period of 1/10th of an epoch. Training one model takes 8 hours on an Nvidia A6000 GPU. For consistency with the original implementations, loss function parameters are $\lambda=25, \mu=25, \nu=1$ for VICReg, and temperature $t=0.5$ for SimCLR. With the encoder's weights frozen, we assess model performance by training a classifier (dense layer with softmax) for 100 epochs using an Adam optimiser and learning rate of $10^{-3}$. We produce supervised baseline comparisons using an equivalent encoder and softmax classifier head, following the same training protocol.

\begin{table}[t]
    \caption{Classification results from linear probing on the NCT dataset evaluation set \cite{kather_dataset} for models trained on the SHIFT dataset \cite{burlingame2020shift}.}
    \label{tab:shift_results}
    \centering
    \begin{tabular}{cccc}
    \toprule
    & & VICReg & SimCLR \\
    Asymmetric & Shared Weights & Accuracy & Accuracy \\
    \midrule
    \checkmark & & $\bm{0.8760}$ & 0.8290 \\
    \checkmark & \checkmark & 0.7934 & 0.8229 \\
    & & 0.8184 & 0.8417 \\
    & \checkmark & 0.8452 & 0.8375 \\
    \bottomrule
    \end{tabular}
\end{table}

\begin{table}[]
    \caption{Classification accuracy for three toy examples for VICReg, SimCLR and supervised baselines. Bold indicates best performance on task for given architecture (VICReg/SimCLR). Tissue/Cell indicates tissue/cell classification task respectively. Tasks use the information-sparse input (Crop/Pad, Downsample, Mask) and model for inference except for the H\&E results, which are paired with the masks for asymmetric architectures.}
    \label{tab:toyresults}
    \centering
    \scalebox{0.8}{\begin{tabular}{lcccccccc}
    \toprule
    Input & & & \includegraphics[width=.07\textwidth]{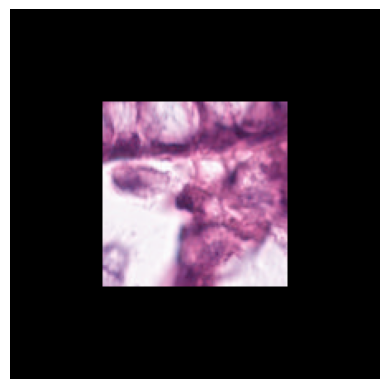} & \includegraphics[width=.07\textwidth]{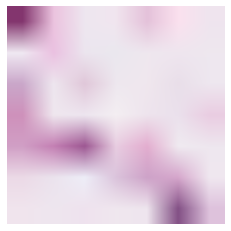} & \multicolumn{2}{c}{\includegraphics[width=.07\textwidth]{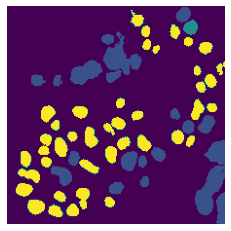}} & \multicolumn{2}{c}{\includegraphics[width=.07\textwidth]{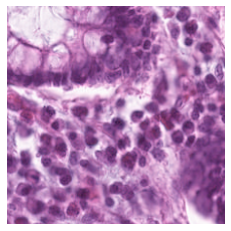}} \\
    & & & Crop/Pad & Downsample & \multicolumn{2}{c}{Mask} & \multicolumn{2}{c}{H\&E} \\
    \midrule
    Model & Asymmetric & Shared Weights & Tissue & Tissue  & Tissue & Cell & Tissue & Cell \\
    \midrule
    \multirow{4}{*}{VICReg} & \checkmark & & 0.8297 & $\bm{0.7743}$ & $\bm{0.5809}$ & $\bm{0.8650}$ & $\bm{0.8979}$ & $\bm{0.8127}$ \\
    & \checkmark & \checkmark & $\bm{0.8334}$ & 0.7730 & 0.3120 & 0.6110 & 0.8725 & 0.7312 \\
    & & & 0.7369 & 0.7383 & 0.5249 & 0.8051 & 0.8419 & 0.6682 \\
    & & \checkmark & 0.7845 & 0.6926 & 0.5338 & 0.8158 & 0.8855 & 0.6904 \\
    \midrule
    \multirow{4}{*}{SimCLR} & \checkmark & & 0.8416 & 0.7505 & $\bm{0.5600}$ & $\bm{0.8540}$ & 0.8850 & $\bm{0.8038}$ \\
    & \checkmark & \checkmark & $\bm{0.8499}$ & $\bm{0.7780}$ & 0.3582 & 0.6053 & 0.8824 & 0.7619 \\
    & & & 0.7902 & 0.7243 & 0.5323 & 0.8151 & 0.8687 & 0.6816 \\
    & & \checkmark & 0.8060 & 0.7241 & 0.5454 & 0.8235 & $\bm{0.9075}$ & 0.6914 \\
    \midrule
    Supervised & - & - & 0.8966 & 0.7656 & 0.5909 & 0.9365 & 0.9451 & 0.8205 \\
    Supervised (Transfer) & - & - & - & - & 0.5345 & 0.9103 & 0.9314 & 0.7045 \\
    \bottomrule
    \end{tabular}}
\end{table}

\section{Results}

\subsection{H\&E-IF Distillation}

As shown in Table \ref{tab:shift_results}, we observe that, for VICReg, the classification performance with asymmetric training is significantly better than with symmetric training, while, for SimCLR, asymmetry results in worse performance. It has been demonstrated that VICReg outperforms other self-supervised methods on tasks with different architectures on each branch, as each branch is regularised independently \cite{bardes2021vicreg}, which is not the case with SimCLR. We also demonstrate in Table \ref{tab:unet_results} that the model considerably outperforms models trained by predicting the IF stain directly from the H\&E stain using a U-Net architecture \cite{ronneberger2015unet}. Direct image-image translation retains primarily fine grained features which are not useful for the downstream task, while self-supervision leads to the extraction of more abstract coarse-grained features.

\subsection{Contextual Distillation}

We observe in Table \ref{tab:toyresults} that asymmetric training leads to considerably more accurate classifications than symmetric training, showing that introducing contextual information into patch models improves performance. In contrast to other examples, there appears to be little effect from sharing weights. Note that asymmetric training significantly outperforms symmetric and even supervised training for the downsampled images. Notably, for all examples in Table \ref{tab:toyresults} asymmetry leads to better performance for SimCLR, in contrast to the results for H\&E/IF distillation.

\subsection{Nuclear Segmentation Distillation}

\noindent In all cases, asymmetric models considerably outperform symmetric models at predictions for masks, as being paired with H\&E patches makes detection of patterns relating to different tissue types easier. Although the accuracy of the tissue classification task is lower for the masks than for the images, it is considerably better than random (0.1111). Despite comparable performance to symmetric models at tissue classification on H\&E patches, asymmetric models perform considerably better at cell classification, as being paired with the nuclear mask forces asymmetric models to extract features relevant to nuclei. It appears that the asymmetry causes models to learn more features irrelevant to tissue classification, but which are essential for cell classification. Similarly, supervised models trained on cell classification from H\&E patches transfer well to tissue classification, while the reverse is not true, as we observe a supervision collapse \cite{doersch2020crosstransformers}.

Fig.~\ref{fig:confmats} shows that the model performs fairly accurately on classes which typically contain at least one nucleus, but accurate classifications of most debris, background and adipose are impossible as the masks are simply matrices of zeros.

\subsection{Comparing Layer Output Similarities}

\begin{figure}[t]
\centering
\begin{subfigure}[b]{0.35\textwidth}
    \centering
    \includegraphics[width=\textwidth]{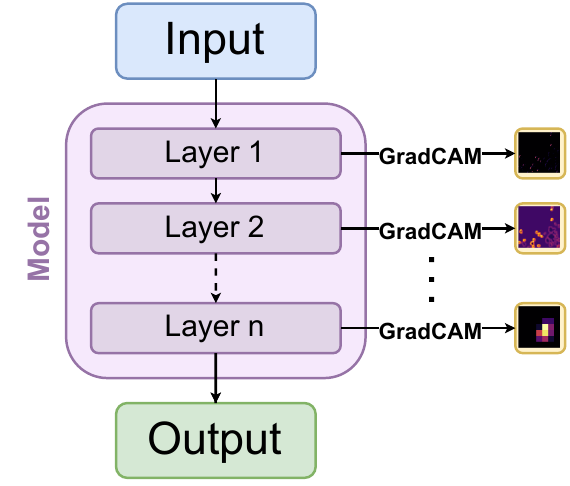}
    \caption{Model}
    \label{fig:gradcamschematic}
\end{subfigure}
\begin{subfigure}[b]{0.29\textwidth}
    \centering
    \includegraphics[width=\textwidth]{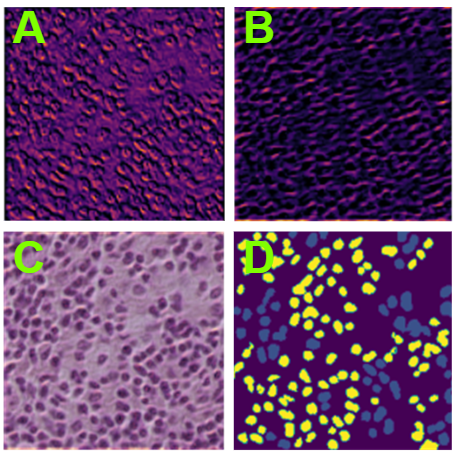}
    \caption{H\&E}
    \label{fig:gradcamhe}
\end{subfigure}
\begin{subfigure}[b]{0.292\textwidth}
    \centering
    \includegraphics[width=\textwidth, trim={0 5 0 0}, clip]{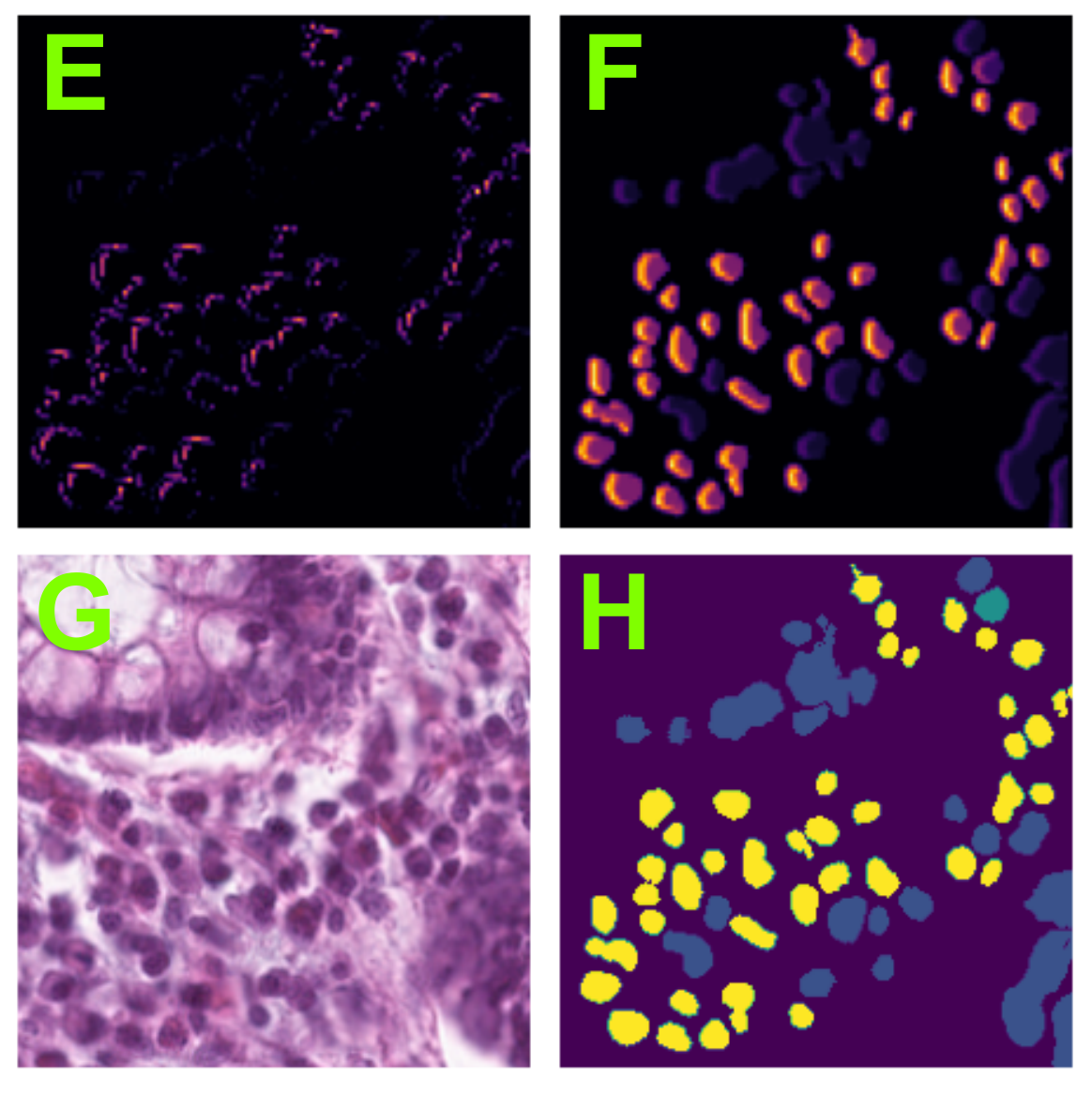}
    \caption{Mask}
    \label{fig:gradcammask}
\end{subfigure}
\caption{(a) Schematic showing where in the model GradCAM images are obtained from, with samples generated from a (b) H\&E patch, and (c) nuclear mask. (A,E) asymmetric GradCAM images, (B,F) symmetric GradCAM images, (C,G) original tissue patches, (D,H) nuclear segmentation masks.}
\label{fig:gradcamfigure}
\end{figure}

\begin{figure}[t]
\begin{subfigure}{0.24\textwidth}
    \centering
    \includegraphics[width=\textwidth]{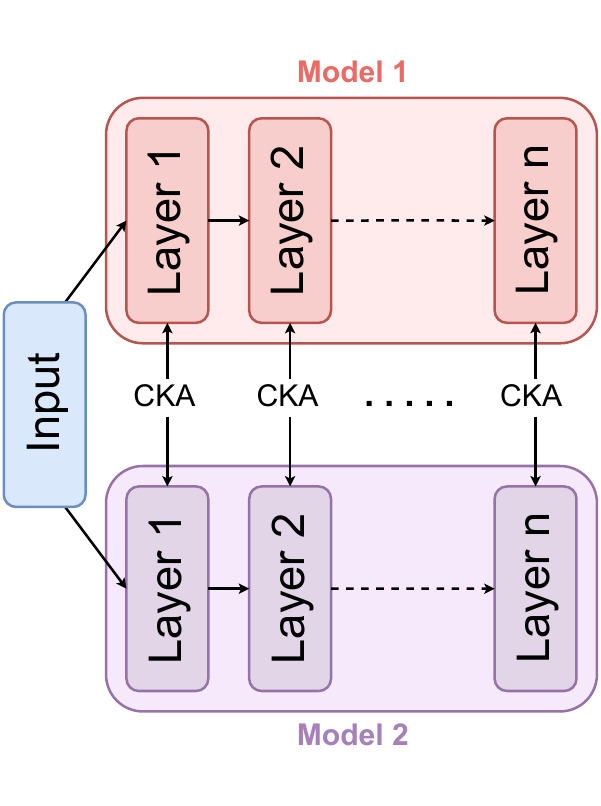}
    \caption{CKA}
    \label{fig:ckaschematic}
\end{subfigure}
\begin{subfigure}{0.7\textwidth}
    \centering
    \includegraphics[width=\textwidth]{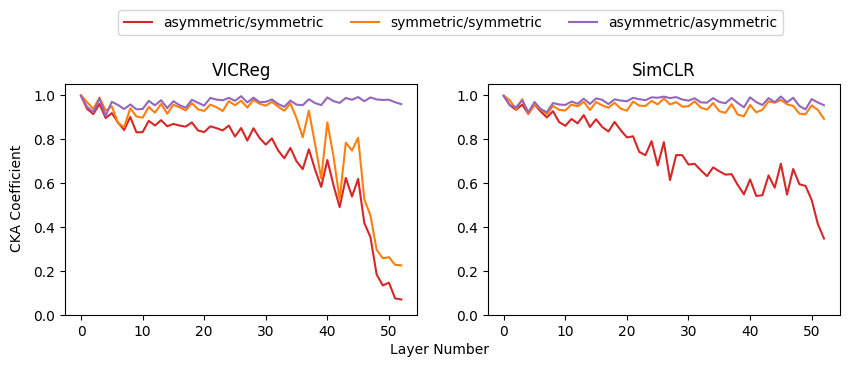}
    \caption{}
    \label{fig:ckaplots}
\end{subfigure}
    \caption{(a) Schematic showing the relationships between layers where centered kernel alignment (CKA) is calculated, (b) CKA analysis of internal representations for H\&E/IF distillation, averaged over pairs from 10 asymmetric and 10 symmetric models. Higher values indicate more similarity between layer outputs.}
    \label{fig:cka}
\end{figure}

To further investigate the difference among learned representations, we use GradCAM \cite{selvaraju2017grad} and centered kernel alignment (CKA) \cite{kornblith2019similarity} to analyse the outputs of each convolutional layer. The differences between layers are demonstrated by the qualitative GradCAM analysis in Fig.~\ref{fig:gradcamfigure}. For H\&E patches (Fig.~\ref{fig:gradcamhe}), GradCAM shows that the nuclei almost always black, meaning the symmetric model ignores sub-nuclear features such as nucleoli. This could explain the lower performance at cell classification, as these areas are heavily focused on by the asymmetric model. For the masks (Fig.~\ref{fig:gradcammask}), symmetrically trained models focus more on coarse-grained morphological features, while asymmetrically trained models focus more on fine-grained features. Fig.~\ref{fig:gradcamlayers} shows that this difference is seen throughout all layers of the model.

We next use CKA, a measure of the similarity of the representations, to show that asymmetric, symmetric and supervised models obtain considerably different representations. Figs.~\ref{fig:cka} and \ref{fig:cka_full} show that asymmetric and symmetric models are highly dissimilar to each other, with the differences increasing throughout the model, and resulting in substantially different representations. The similarity score between asymmetric/asymmetric pairs is high, while similarity between symmetric/symmetric pairs is high only for SimCLR. This suggests that, with VICReg, there is a larger feature space which can be learned without the restriction imposed by asymmetry, leading to more variability between symmetric models.

\section{Discussion and Conclusions}

Our results show that asymmetric training can improve the performance of models on downstream tasks by extracting more relevant features from information-sparse inputs. Each branch in a joint-embedding architecture utilises the output of the other branch as a supervisory signal, meaning both branches extract approximately the same features from their input. The asymmetric pairing forces the information-sparse encoder to find patterns in the data which correlate with those found in the information-dense images, while possibly disregarding patterns which are easier to detect. This means that subtler features can be found which are harder to detect in the information-sparse input. In the symmetric case, the model is less able to focus on these weakly-signalled features, as they may not be as easily distinguished from noise.

Despite the objective forcing embeddings to be as similar as possible, we still observe significant differences in the classification accuracy of each branch, particularly in the nuclear mask example. It has been shown that the use of projection heads improves representation quality \cite{chen2020simple,mialon2022variance}; we conjecture that the projection head filters out information irrelevant to the opposite branch. This would explain the high classification accuracy observed in the asymmetrically trained H\&E encoders for the nuclear mask example, which remains comparable with symmetric original image encoders despite the significantly lower performance of the paired mask encoders.

\subsection{Asymmetric Learning Obtains Measurably Different Representations}

In asymmetric models, we observe both qualitative and quantitative differences in the representations of both information-dense and information-sparse inputs compared to symmetric models. Intuitively, this is because knowledge distillation is bidirectional, so while knowledge is distilled from the information-dense model into the information-sparse model, there is also knowledge distilled from the information-sparse model to the information-dense model.

Our CKA analysis, shown in Figs.~\ref{fig:cka} and \ref{fig:cka_full}, further quantifies the difference between the layers of symmetric, asymmetric and supervised models. Models with different initialisations become very similar to other models with the same training regime, but significantly different to models with a different training regime, for both information-sparse and -dense data. We conclude that the representations of asymmetrically trained models contain different features from those of symmetrically trained models. This is corroborated by our qualitative GradCAM analysis in Fig.~\ref{fig:gradcamfigure}, where we demonstrate that the choice of paired data has a significant effect on the areas of the image which are focused on by the model. This could be leveraged to train models to target desirable features such as nucleoli without the arduous process of manual labelling. Utilising asymmetric training approaches, models could be developed for evaluating routine data, such as processing H\&E stains, without the need for costly additional staining.

There are a plethora of histological datasets containing different modalities, inputs, or views, and asymmetry can be leveraged to detect features in routine data which are otherwise detectable only in data that is harder to obtain. Architectures could also utilise aspects of typical knowledge distillation frameworks such as having a larger \emph{teacher} network for the paired data to ensure more informative representations are obtained from the information-dense inputs, improving the efficacy of the technique. 

\subsubsection{Acknowledgements} LF is supported by the MRC grant MR/W006804/1, RI is supported by EPSRC grant EP/S0300875/1 and Wellcome grant 221786/Z/20/Z. KY acknowledges support from EP/R018634/1, BB/V016067/1, and European Union’s Horizon 2020 research and innovation programme under grant agreement No 101016851, project PANCAIM. We thank Dr. Adalberto Claudio-Quiros for his helpful feedback and support.

\bibliographystyle{splncs04}
\bibliography{bib}

\newpage

\section*{Supplementary Information}

\newcommand{\beginsupplement}{
    \setcounter{section}{0}
    \renewcommand{\thesection}{S\arabic{section}}
    \setcounter{equation}{0}
    \renewcommand{\theequation}{S\arabic{equation}}
    \setcounter{table}{0}
    \renewcommand{\thetable}{S\arabic{table}}
    \setcounter{figure}{0}
    \renewcommand{\thefigure}{S\arabic{figure}}
    \newcounter{SIfig}
    \renewcommand{\theSIfig}{S\arabic{SIfig}}}

\beginsupplement

\begin{table}[]
    \caption{Augmentations and probability of application. Augmentations are randomly applied at each training iteration, with the given probability.}
    \refstepcounter{SIfig}\label{tab:augmentations}
    \centering
    \scalebox{1}{\begin{tabular}{lc}
    \toprule
    Augmentation & Probability\\
    \midrule
    Flip (left/right/up/down) & 1.0 \\
    Crop (scale 75\% to 100\%) & 1.0 \\
    Gaussian Background Noise & 0.3\\
    Rotation & 0.4 \\
    Solarize & 0.3 \\
    Colour Jitter & 1.0\\
    \bottomrule
    \end{tabular}}
\end{table}

\begin{table}[]
    \centering
    \caption{Label distributions for the train and test splits in the NCT dataset. For the train set we use "NCT-CRC-HE-100K" and for the test set we use the external validation set "CRC-VAL-HE-7K", with no overlap of patients between the two sets, as detailed at \cite{kather_dataset}. Tissue classification labels are abbreviated as follows: Adipose (ADI), background (BACK), debris (DEB), lymphocytes (LYM), mucus (MUC), smooth muscle (MUS), normal colon mucosa (NORM), cancer-associated stroma (STR), colorectal adenocarcinoma epithelium (TUM)}
    \label{tab:class_balance}
    \scalebox{0.72}{\begin{tabular}{llccccccccc}
        \toprule
        Task & Split & \multicolumn{9}{c}{Labels} \\
        \midrule
        & & TUM & STR & NORM & MUS & MUC & LYM & DEB & BACK & ADI \\
        \midrule
        \multirow{2}{*}{Tissue Classification} & Train & 14311 & 10444 & 8760 & 13534 & 8892 & 11556 & 11503 & 10564 & 10404 \\
        & Test & 1232 & 421 & 741 & 592 & 1034 & 634 & 339 & 847 & 1337 \\
        \midrule
        & & \multirow{2}{*}{Background} & \multirow{2}{*}{Neoplastic} & \multirow{2}{*}{Inflammatory} & \multirow{2}{*}{Connective} & \multirow{2}{*}{Dead} & Non-Neoplastic \\
        & & & & & & & Epithelial \\
        \midrule
        \multirow{2}{*}{Cell Classification} & Train & 19028 & 34891 & 3357 & 22229 & 18495 & 1968 \\
        & Test & 1613 & 2501 & 415 & 1540 & 1054 & 54 \\
        \bottomrule
    \end{tabular}}
\end{table}

\begin{table}
\centering
\caption{Effect of dataset size on classification performance. Models were trained on the given fraction of the training dataset for both self-supervised training and classifier head training, and then evaluated on the 100\% of validation data.}
\label{tab:dataset_size}
\begin{tabular}{cccccc|cccc}
\toprule
    & & \multicolumn{4}{c}{VICReg} & \multicolumn{4}{c}{SimCLR} \\
    & & \multicolumn{2}{c}{H\&E} & \multicolumn{2}{c}{Mask} & \multicolumn{2}{c}{H\&E} & \multicolumn{2}{c}{Mask} \\
    Fraction & Asymmetric &          Tissue &      Cell &        Tissue &      Cell &  Tissue &      Cell &        Tissue &      Cell \\
\midrule
\multirow{2}{*}{0.01} & \checkmark &        0.4349 &  0.4602 &      \textbf{0.2732} &  \textbf{0.5412} &        0.6011 &  \textbf{0.5376} &      \textbf{0.3882} &  \textbf{0.6407} \\
    &  &        \textbf{0.6041} &  \textbf{0.5281} &      0.1863 &  0.3485 &        \textbf{0.6149} &  0.5116 &      0.1863 &  0.3485 \\
\midrule
\multirow{2}{*}{0.02} & \checkmark &        0.5339 &  0.4945 &      \textbf{0.3630} &  \textbf{0.6642} &        0.6210 &  \textbf{0.5782} &      \textbf{0.4000} &  \textbf{0.6942} \\
    &  &       \textbf{0.6819} &  \textbf{0.5218} &      0.3281 &  0.3485 &        \textbf{0.6421} &  0.5238 &      0.1877 &  0.3485\\
\midrule
\multirow{2}{*}{0.05} & \checkmark &        0.6483 &  0.5617 &      \textbf{0.4116} &  \textbf{0.6813} &        0.7357 &  \textbf{0.6613} &      \textbf{0.4464} &  \textbf{0.7070} \\
    &  &        \textbf{0.7730} &  \textbf{0.5739} &      0.1654 &  0.4868 &        \textbf{0.7829} &  0.5629 &      0.2292 &  0.5474 \\
\midrule
\multirow{2}{*}{0.1} & \checkmark &        0.7747 &  \textbf{0.6302} &      \textbf{0.4905} &  \textbf{0.7305} &        0.8223 &  \textbf{0.7063} &      \textbf{0.4827} &  \textbf{0.7506} \\
    &  &        \textbf{0.8402} &  0.6015 &      0.3836 &  0.5467 &        \textbf{0.8767} &  0.6179 &      0.3560 &  0.5310 \\
\midrule
\multirow{2}{*}{0.2} & \checkmark &        0.8432 &  \textbf{0.7477} &      \textbf{0.5150} &  \textbf{0.8254} &        0.8641 &  \textbf{0.7716} &      \textbf{0.5082} &  \textbf{0.8318} \\
    &  &        \textbf{0.8778} &  0.6375 &      0.3651 &  0.6696 &        \textbf{0.8923} &  0.6404 &      0.3726 &  0.5965 \\
\midrule
\multirow{2}{*}{0.5} & \checkmark &        0.8509 &  \textbf{0.7847} &      \textbf{0.5349} &  \textbf{0.8437} &        \textbf{0.9096} &  \textbf{0.7861} &      \textbf{0.5356} &  \textbf{0.8509} \\
    &  &        \textbf{0.8754} &  0.6745 &      0.4159 &  0.7137 &        0.9027 &  0.6943 &      0.3960 &  0.7040 \\
\midrule
\multirow{2}{*}{1} & \checkmark &        \textbf{0.8979} &  \textbf{0.8127} &      \textbf{0.5809} &  \textbf{0.8650} & 0.8850 & \textbf{0.8038} & \textbf{0.5600} & \textbf{0.8540} \\
    &  &        0.8855 &  0.6904 &      0.5338 &  0.8158 & \textbf{0.9075} & 0.6914 & 0.5454 & 0.8235 \\
\bottomrule
\end{tabular}
\end{table}

\begin{table}
    \caption{Encoder Ablations. Results are found to be robust to the choice of encoder, with ResNet50 \cite{he2016deep}, MobileNetV2 \cite{howard2017mobilenets}, EfficientNetB0 \cite{tan2021efficientnetv2} and Xception \cite{chollet2017xception} tested. Bold type indicates best self-supervised performance on downsampled image predictions. Information-dense results (Full Resolution) are included to show the negative effect on the performance of the information-dense branch unless the task is highly specific, as shown in Table \ref{tab:toyresults}.}
    \label{fig:encablations}
    \centering
     \scalebox{1}{\begin{tabular}{lccc|ccc}
     \toprule
     \multirow{2}{*}{Encoder} & \multicolumn{3}{c}{Downsampled} & \multicolumn{3}{c}{Full Resolution}\\
     & Asymmetric & Symmetric & Supervised & Asymmetric & Symmetric & Supervised\\
     \midrule
     ResNet50 & $\bm{0.7743}$ & 0.6926 & 0.7656 & 0.7890 & 0.8846 & 0.9370 \\
     MobileNetV2 & $\bm{0.7580}$ & 0.7050 & 0.7616 & 0.7959 & 0.8602 & 0.9390 \\
     EfficientNetB0 & $\bm{0.7757}$ & 0.7109 & 0.7595 & 0.8265 & 0.8441 & 0.9241 \\
     Xception & $\bm{0.7619}$ & 0.6858 & 0.7719 & 0.8258 & 0.8965 & 0.9351 \\
     \bottomrule
    \end{tabular}}
\end{table}

\begin{table}[]
    \centering
    \caption{Results for using U-Net \cite{ronneberger2015unet} with same training parameters as Fig.~\ref{tab:shift_results}, including the ResNet-50 backbone, to generate the IF images directly from the H\&E images in the SHIFT dataset \cite{burlingame2020shift}. Loss functions used were structural similarity (SSIM) and mean squared error (MSE), and the effect of using augmentations on the training data was also considered. A classifier head was then trained with the ResNet-50 backbone as above, with accuracy determined by tissue type classification on the NCT dataset \cite{kather_dataset}. Comparable accuracies for self-supervised models are shown in Table \ref{tab:shift_results}.}
    \label{tab:unet_results}
    \begin{tabular}{llcc}
    \toprule     
        Model & Loss & Augmentations & Accuracy \\
    \midrule
        \multirow{4}{*}{U-Net} & SSIM & \checkmark & 0.4319 \\
        & SSIM &  & 0.5646 \\
        & MSE & \checkmark & 0.5975 \\
        & MSE &  & 0.6784 \\
        \midrule
        VICReg (Symmetric) & & \checkmark & 0.8452 \\
        VICReg (Asymmetric) & & \checkmark & \textbf{0.8760} \\
    \bottomrule
    \end{tabular}
\end{table}

\begin{table}[]
    \centering
    \caption{Results for model trained on the NCT dataset paired with HoVer-Net masks, and evaluated on classifying the CoNIC dataset \cite{graham2021lizard,graham2021conic} by the predominant cell type in each image. Note that the classes (Epithelial, Lymphocyte, Plasma, Neutrophil, Eosinophil, Connective) being classified are different to the task in Table \ref{tab:toyresults}, so results are not directly comparable. This is particularly relevant for the \emph{mask} results, as the SSL training masks (NCT) and classification training masks (CoNIC) contain different nuclei class labels.}
    \label{tab:conic_results}
    \begin{tabular}{ccc}
    \toprule
         Asymmetric & H\&E Accuracy & Mask Accuracy \\
    \midrule
        \checkmark & 0.8397 & 0.8537 \\
        & 0.7816 & 0.8437 \\
    \bottomrule
    \end{tabular}
\end{table}

\begin{figure}[b]
    \centering
    \begin{subfigure}{0.45\textwidth}
        \centering
        \includegraphics[width=\textwidth]{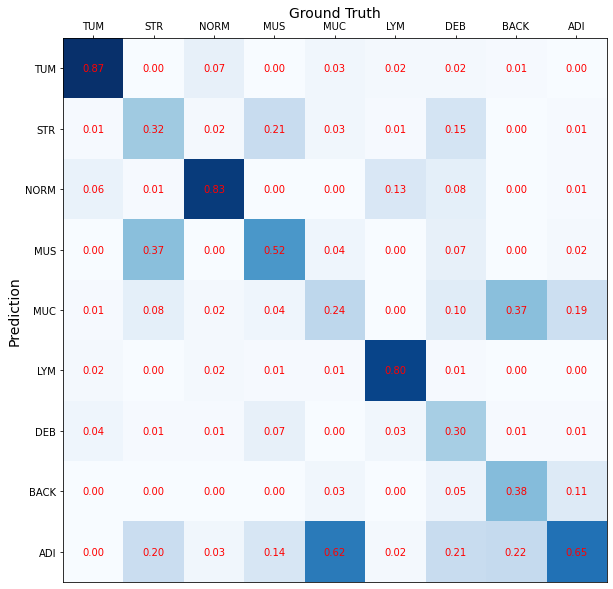}
        \caption{Asymmetric}
        \label{fig:asymmconfmat}
    \end{subfigure}
    ~
    \begin{subfigure}{0.45\textwidth}
        \centering
        \includegraphics[width=\textwidth]{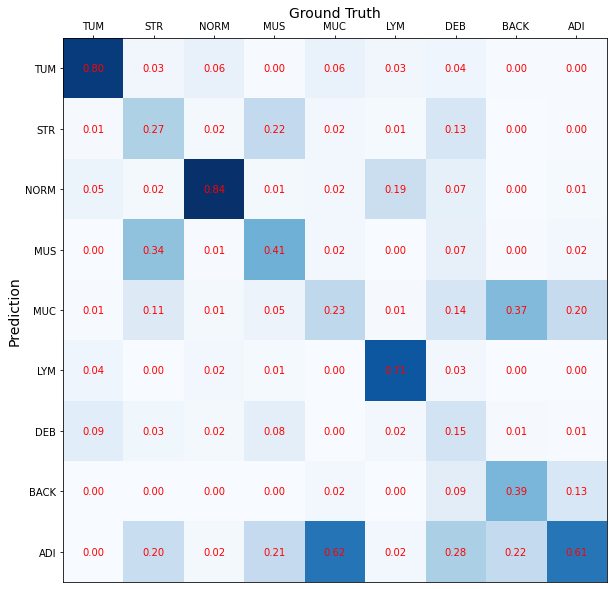}
        \caption{Symmetric}
        \label{fig:symmconfmat}
    \end{subfigure}
    \caption{Confusion Matrices for mask predictions of classifier trained on (a) image/mask and (b) mask/mask pairs.}
    \label{fig:confmats}
\end{figure}

\begin{figure}
    \centering
    \begin{subfigure}{0.45\textwidth}
    \includegraphics[width=\textwidth]{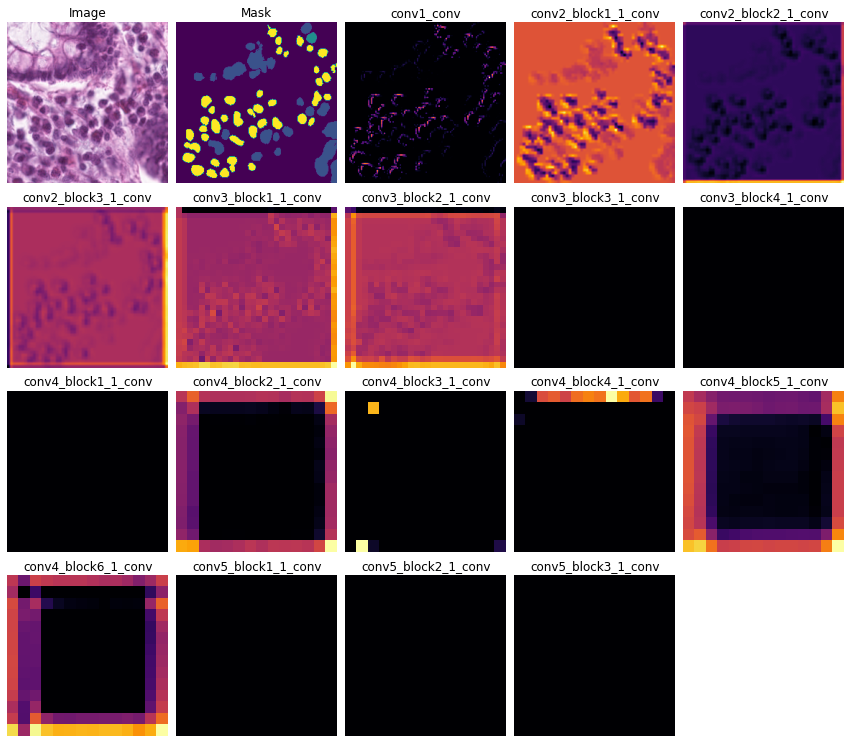}
    \caption{Asymmetric}
    \label{fig:asymmlayers}
    \end{subfigure}
    ~
    \begin{subfigure}{0.45\textwidth}
    \includegraphics[width=\textwidth]{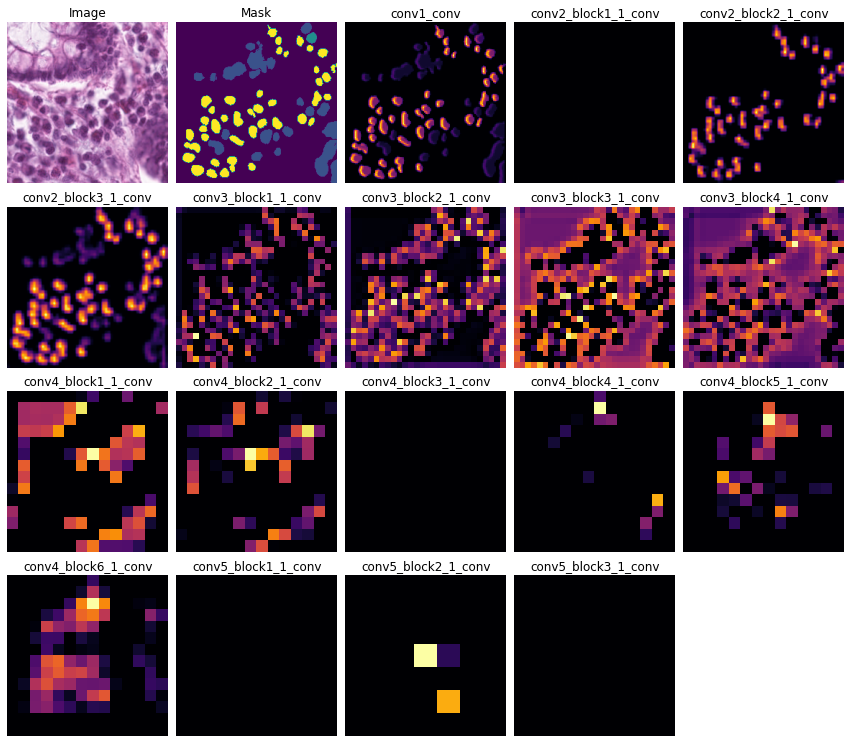}
    \caption{Symmetric}
    \label{fig:symmlayers}
    \end{subfigure}
    \caption{GradCAM analysis of first convolutional layer from each ResNet block in model trained on (a) image/mask and (b) mask/mask pairs.}
    \label{fig:gradcamlayers}
\end{figure}

\begin{figure}
    \centering
    \includegraphics[width=\textwidth]{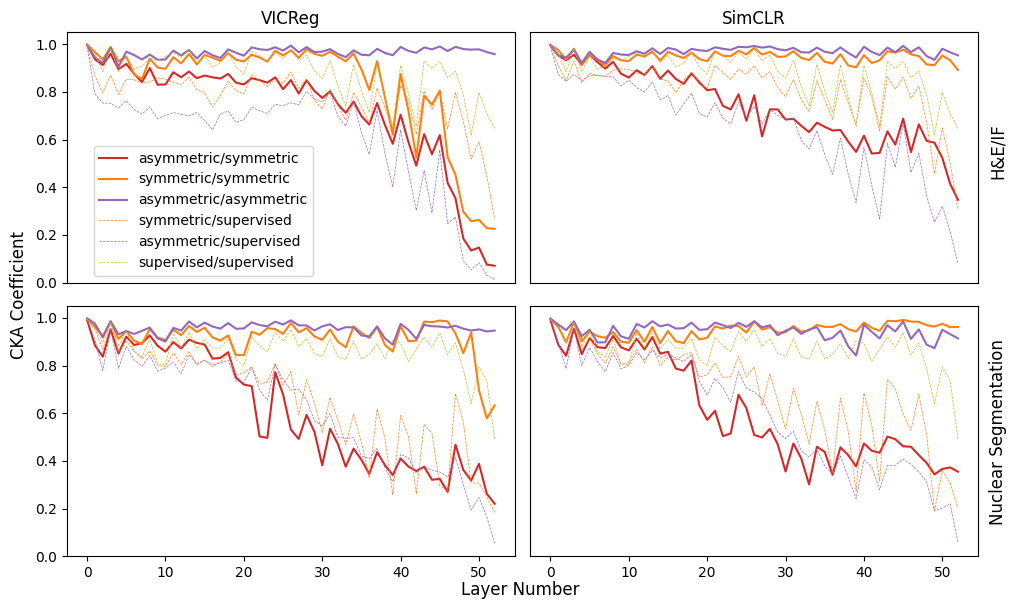}
    \caption{CKA analysis of internal representations for H\&E/IF distillation and Nuclear Segmentation Distillation, averaged over pairs from 10 asymmetric, 10 symmetric and 10 supervised models. Higher values indicate more similarity between layer outputs.}
    \label{fig:cka_full}
\end{figure}

\end{document}